# Operation of a THGEM-based detector in low-pressure Helium

**Marco Cortesi,**[*] **John Yurkon, and Andreas Stolz**

*National Superconducting Cyclotron Laboratory, Michigan State University*
*East Lansing, Michigan 48824, USA*
E-mail: cortesi@nscl.msu.edu

ABSTRACT: In view of a possible application as a charge-particle track readout for an Active Target Time Projection Chamber (AT-TPC), the operating properties of THick Gaseous Electron Multipliers (THGEM) in pure low-pressure Helium were investigated. This paper includes the effective gain dependence on pressure for different detector configurations (single-, double-, triple-cascade setup), long-term gain stability and energy resolution from tracks of 5.5 MeV alpha particles. Stable operational conditions and maximum detector gains of $10^4$-$10^7$ have been achieved in pure Helium at pressure ranging from 100 torr up to 760 torr. Energy resolution of 6.65% (FWHM) for 690 keV of energy deposited by 5.5 MeV alpha particles at 350 torr was measured. The expected energy resolution for the full track is around 2.4% (FWHM). These results, together with the robustness of THGEM electrodes against spark damage, make THGEM structures highly competitive compared to other technologies considered for TPC applications in an active target operating with pure noble gases, requiring a high dynamic range and a wide operating pressure range down to few hundred torr.

---

[*] Corresponding author.

# Contents

## 1. Introduction

In the field of experimental nuclear physics and astrophysics, a number of significant improvements in production of radioactive isotope beams (RIBs) have been recently achieved. This opens the possibility to probe nuclei that are far away from the valley of stability by means of relatively simple and well understood reactions, such as elastic and inelastic scattering, one or two nucleon transfer, or Coulomb excitation.

Active-Target Time Projection Chambers (AT-TPC), in which a large volume of a low-mass filling gas acts simultaneously as target and ionization medium, are one of the most efficient approaches for implementing RIB experiments. They are extremely versatile, providing a thick target without loss of energy resolution, a $4\pi$ solid angle coverage, and a large dynamic range which can be achieved by adjusting the pressure of the target. To simplify the analysis of experimental observables very pure low-mass gases with simple structure (such as Hydrogen, Deuterium, Helium-3 and Helium-4) are preferred. A pure active target TPC allows unambiguous event-by-event kinematic reconstruction for identifying particular reaction channels or identifying exotic forms of radioactive decay, permitting a powerful and incisive method to elucidate spectroscopic and structural characteristics of exotic nuclei.

Presently, the most promising position-sensitive charge readout technologies for AT-TPC applications are based on Micro-Pattern Gaseous Detectors (MPGD) [1]. These new devices achieved remarkable improvements compared to traditional wire-based chambers, including a better spatial resolution due to the smaller intrinsic size of the gas amplification volumes, a more stable and simplified operations, as well as a higher counting rate capabilities. Hole-type multipliers, such as Gaseous Electron Multiplier (GEM) [2], and Micro-Mesh Gaseous Structure (Micromegas) [3] are nowadays well-established detector technologies.

The development of a new generation of MPGD-readout AT-TPC detectors are currently pursued by several groups worldwide. Successful operation of a first prototype AT-TPC (PAT-ATC), based on Micromegas, has been recently reported [4], [5] by the National Superconducting Cyclotron Laboratory (NSCL) at Michigan State University (MSU). A larger version of this AT-TPC has been recently commissioned and it will soon be fully operational. Among others,

ACTAR TPC [6] and GEM-MSTPC [7], [8] are two examples of GEM/Thick GEM-based AT-TPCs currently under investigation.

In order to suppress avalanche-feedback processes and to provide a stable operational conditions, MPGD-based TPCs generally operate in noble gas mixtures (typically He-based mixtures) with a small percentage of polyatomic (hydrocarbon) compounds characterized by a high quenching efficiency. The MPGD-based TPCs mentioned above were operating mainly in He-based mixtures (He-$CO_2$, He-$CF_4$, etc.). However, in contrast to open-geometry structures such as wire-based detectors or Micromegas, high-gain operation is attainable with hole-type structures, where instabilities initiated by secondary (non-quenched) photon-feedback effects are prevented by the strong confinement of the avalanche within sub-millimeter holes. High-gain operation of GEM- and thick GEM-based detectors has been demonstrated in pure noble gases at atmospheric and higher pressures by various groups [9]–[13]. In addition, several of these groups report that some impurities, from outgassing or residual gas, may play a critical role on the operation of GEM-like detectors in noble gases. In particular, it has been shown that Nitrogen, most probably due to ion charge exchange [11] as well as the Penning effect [14], allows for achieving high electron avalanche multiplication at relative low voltage. Most recently, effective gas gain factors above $10^3$ were also attained with a thick GEM-based detector in pure deuterium at low pressure [15].

The present study encompasses a detailed evaluation of the operational conditions of Thick GEM-like (THGEM) detector operated with low-pressure helium for the possible application as a position-sensitive AT-TPC charge readout. The THGEM [16]–[18] is a robust hole-type structure, resembling that of a standard GEM but with 5–20-fold larger dimensions, suitable for operation at very low pressure [19]. Such structures can be manufactured economically by standard printed-circuit drilling and etching technology, while material, parameters and shape can be tailored to the specific application.

The interest in helium as an active target for inverse kinematic experiment with RIBs relies on the crucial role played by scattering and transfer reaction measurements by alpha particles, needed to address important open questions in nuclear structure and astrophysics [20], [21]. Detailed measurements performed in a large range of pressure, from 100 torr to 760 torr, and with various detector configurations (single-, double- and triple-cascade THGEM detector) will be presented; in particular we discuss effective gains, long-term gain stability, and energy resolution from energy deposited by 5.5 MeV alpha particles, as well as future prospects.

## 2. Detector setup and methods

Figure 1 illustrates the various THGEM-based detector configurations used for effective gain measurements; they comprise either a single WELL-THGEM [22], [23], a double- or a triple-cascade elements with the last electrode in WELL configuration. Both, the single- and double-sided copper-clad THGEMs used throughout the present work have an effective area of 100 x 100 mm$^2$, a thickness of 0.6 mm, a hole pitch of 0.7 mm, a hole diameter of 0.3 mm, and a rim around the holes of 0.1 mm.

Due to the long range of electrons in He (e.g. photocapture electrons from 5.9 keV X-rays have a continuous-slowing-down approximation range of around 5 mm at standard ambient temperature and pressure), measuring gain curves by a linear fit to the energy/pulse-height relation (pulse mode) is impractical. This is particularly true at low pressure where poor energy resolution, as a result of a substantial edge effect, makes it impossible to distinguish the photo-peak. To overcome this problem, measurements of effective gain curves have been carried out in current mode with a methodology similar to the one presented in [17]. The top surface of the first THGEM was illuminated with UV light from a continuous Ar(Hg) lamp. The current signals generated by the avalanche multiplication process were recorded on the readout electrodes, which were grounded through a high-precision electrometer (Keithley Model 614) – see Figure 1. The detector gain was conventionally defined as the anode current ($I_{OUT}$) divided by the current induced by the UV-light on the THGEM top surface ($I_{IN}$). The latter was determined in a dedicated set of measurements (Figure 2a), in which a reverse electric field was applied on the drift gap and the photocurrent was recorded on the cathode mesh. $I_{IN}$ was defined as the current collected on the cathode at the middle of the ion-chamber region of the current-voltage plot, between the steeply rising recombination region and the onset of proportional multiplication. Figure 2b depicts the photocurrents as function of the drift field for various filling gas pressures.

The maximum achievable gain was defined when detector discharges reached a threshold of a few sparks per minute; before the maximum achievable gain was attained, the probability of the detector discharges was negligible (no discharges have been observed during the few hours of data taking in each measurement).

Each electrode was individually biased through a 5 MΩ resistor using a high-voltage power supply (Tennelec TC 953). This allows an adjustable electric field strength to be applied across and between the various detector elements. The electric field between THGEMs (transfer field) was usually set at 250 V/cm, while equal voltage bias was applied to the top surface of the first cascade THGEM (acting as the photocathode) and the cathode mesh as a zero drift field provides maximum electron collection efficiency in the case of a reflective photocathode arrangement [17].

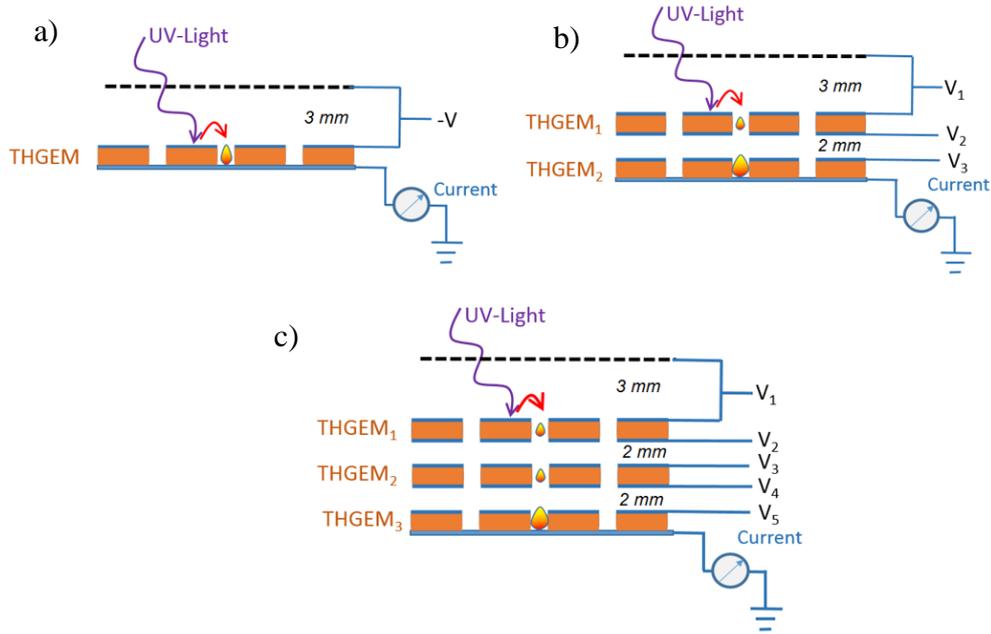

**Figure 1. Setups of the THGEM-based detector prototypes used for effective gain measurement: a) single THGEM in WELL configuration, b) and c) double- and triple- cascade elements, with the last element in WELL configuration.**

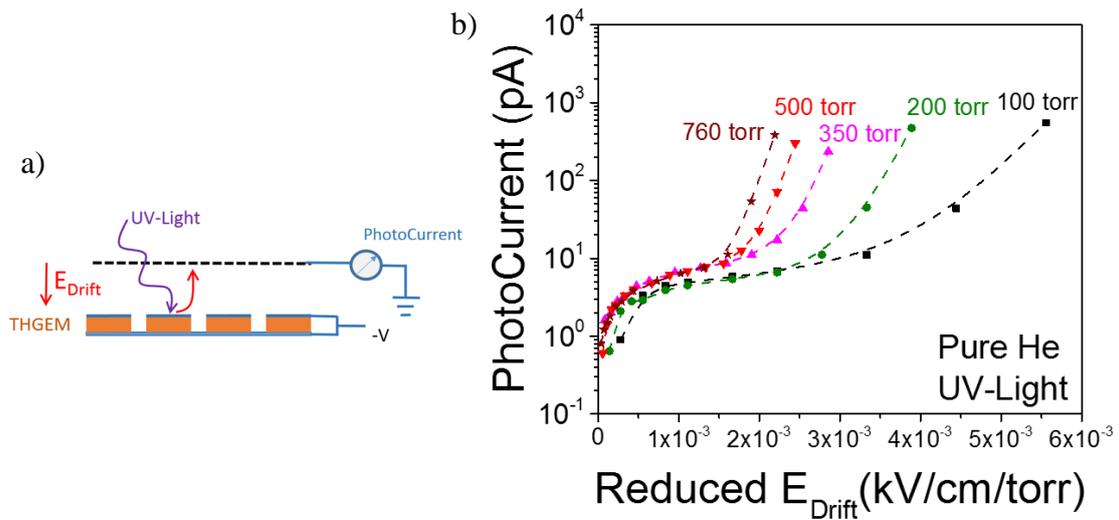

**Figure 2.** Part a) Schematic drawing of the setup for the measurements of the photocurrent induced by the UV-light illumination on the top-surface of the first cascade THGEM element. Part b) Photocurrent as function of the reduced voltage applied to the drift gap, for the relevant pressure values.

The pressure inside the detector vessel was regulated by a MKS type 146A multiple-purpose control unit and by a differential pressure gauge (MKS Baratron Type 223B), while the gas flow (in the range of 5-30 sccm) was set by a flowmeter mounted along the vacuum-pump line. A schematic drawing of the gas handling system is presented in Figure 3.

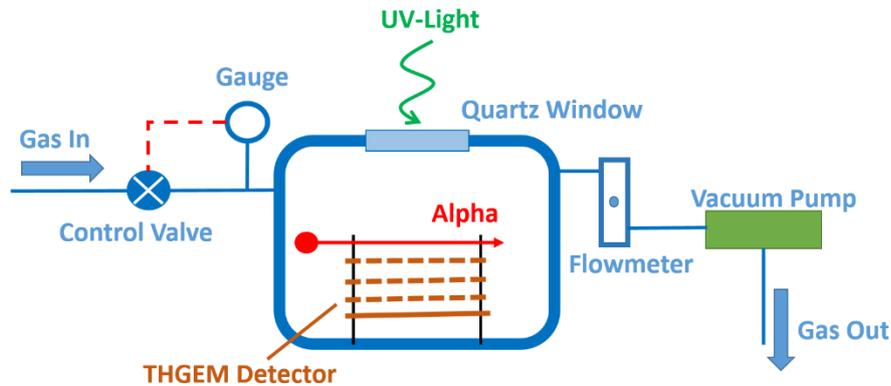

**Figure 3.** Schematic drawing of the gas handling system.

Prior to any measurements, the detector vessel was evacuated with a dual stage rotary vane vacuum pump (Pfeiffer Balzers Duo 008B) down to a residual pressure of few mTorr, and then flushed with He for several hours in order to preserve the purity of the gas. The gas was supplied from an ultra-high purity (99.999%) Helium gas cylinder; effects of residual impurities (mainly outgassing from polymer components at a rate below $10^{-4}$ Torr-L/sec in the vessel volume) on effective gain are discussed in section 3.1. Due to an initial fast variation of the effective gain caused by charging up and polarization effects, the data shown in section 3.1 were taken after 10-15 minutes whenever the biases of the THGEM electrode were changed.

The effective gain stability over time was monitored recording the electrometer reading with a programmable microprocessor (Arduino Due), while the detector was illuminated with UV photons for several hours. Results of effective gain at different pressures and long-term gain stability measurement are presented and discussed in section 3.1 and section 3.2 respectively.

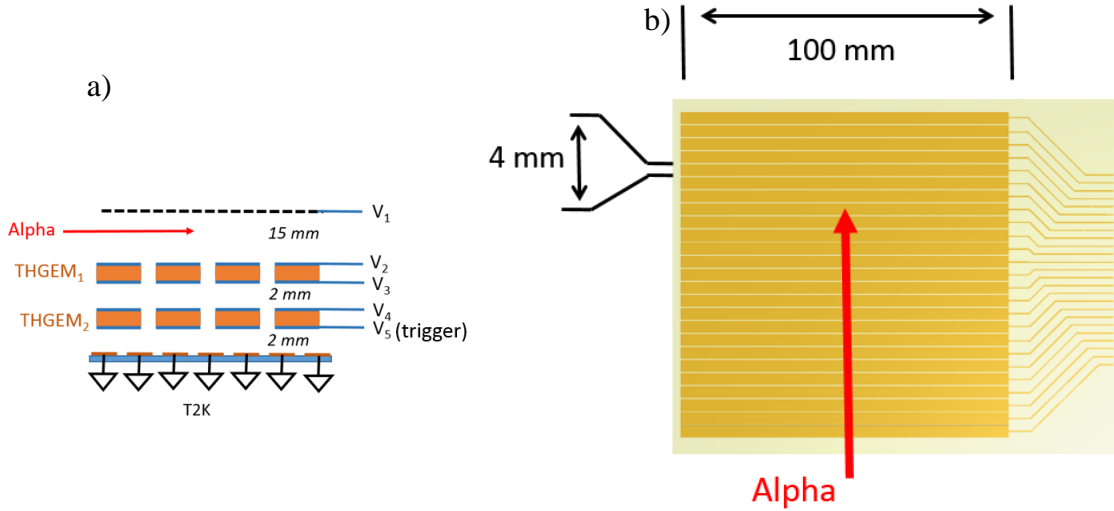

**Figure 4. a)** Schematic drawing of the THGEM-based detector setups used for recording the pulse-height distribution from 5.5 MeV alpha track. **b)** Top view of the strips readout electrode.

Finally, pulse-height signals from 5.5 MeV alpha tracks were acquired with a two-cascade THGEM detector coupled to a strip anode readout (Figure 4). This detector had an effective area of 104.8 x 104.8 mm$^2$, segmented into 25 strips. Each strip was 25 mm wide, while the gap between neighbouring strips was 0.2 mm. The strip signals were routed into connectors mounted at the sides of the readout PCB, and sequentially read and processed by the T2K electronics [24]. The TTL trigger was provided from the bottom electrode of the second cascade THGEM, after the signals had been processed by a charge sensitive preamplifier (Canberra 2003BT) and by a gate and delay generator (ORTEC model 416A). A fine mesh, placed 15 mm (drift gap) atop the first cascade THGEM, functioned as cathode electrode; the transfer gap (distance between the two THGEMs) and the induction gap (distance between the last cascade THGEM and the readout anode) were both 2 mm wide. A collimated Am-241 source provided irradiation with 5.5 MeV alpha particles with tracks perpendicular to the readout strips. The source was placed within the drift gap at a distance of around 70 mm from the detector active area, providing a trigger rate of about 10 Hz. Analysis and results from this test are presented and discussed in section 3.3.

## 3. Results

### 3.1 Effective gain

Figure 5 depicts the avalanche gain as a function of the reduced bias (voltage/pressure) symmetrically applied to each THGEM elements, measured for pressures ranging from 100 torr to 760 torr and for three detector configurations, schematically illustrated in Figure 1.

As a result of the pressure decrease that leads to a longer electron mean free path, the exponential gain-voltage dependence has a different scale according to the filling pressure. It shall be noted that the major feature of the gain curves of Figure 5, common to all the detector configurations, is a systematic decrease of the maximum achievable gain while lowering the pressure. For the high pressure range (500-760 torr), large maximum achievable gain (~$10^5$-$10^7$) is reached at very low reduced voltages (below 1 Volt/torr). At lower pressures (for instance 100 torr), the maximum achievable gain is reduced by three orders of magnitude (~$10^2$-$10^4$) and it is attained at relative large reduced bias (> 2.5 Volt/torr).

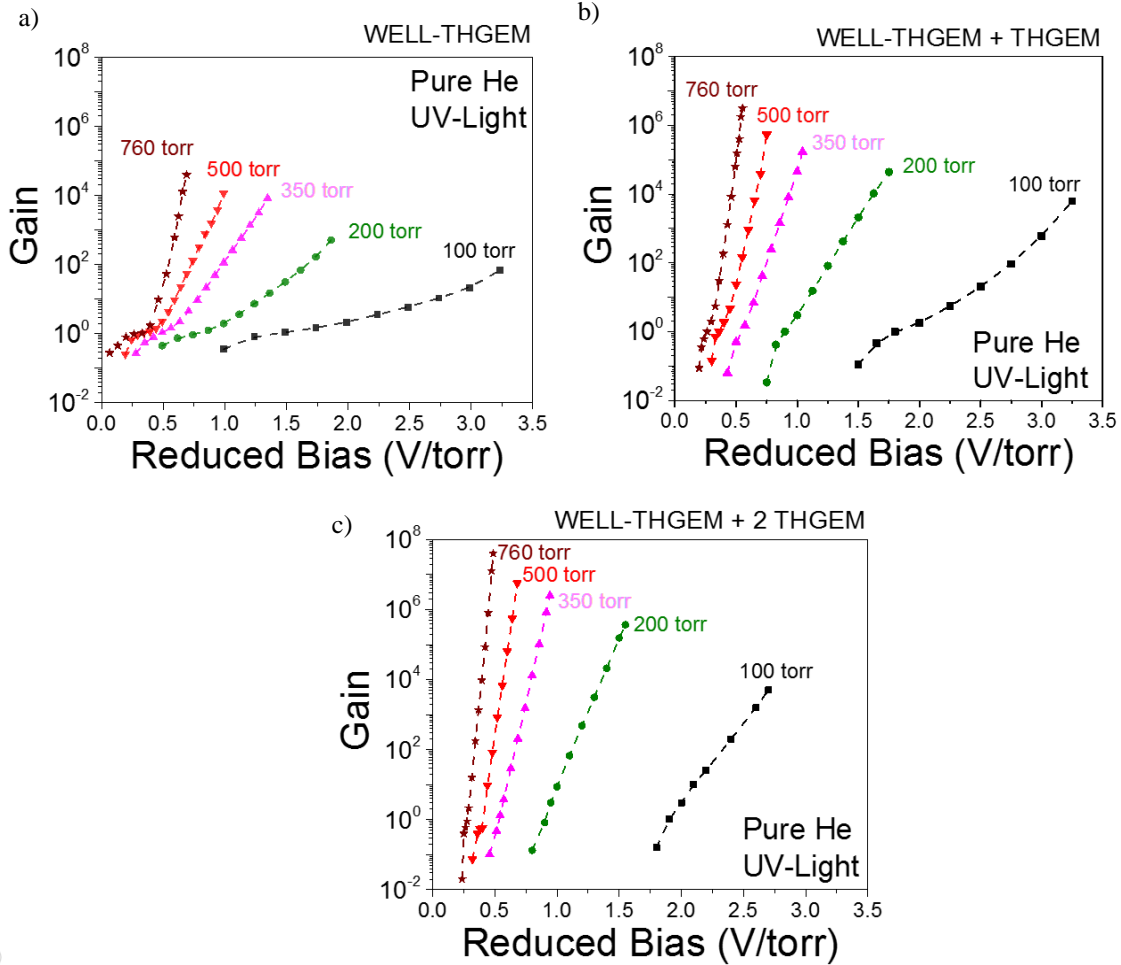

**Figure 5. Effective gain curves in the range of 100-760 torr for the three different THGEM-based detector configurations depicted in Figure 1.**

In order to investigate the validity of the pressure scaling of the detector gain, the data set obtained for each detector configuration and at different pressures has been further analyzed. All gain curves of Figure 5 have been fitted with an exponential function model

$$G \sim \exp(\gamma\, V/p) \qquad \text{Eq. 1}$$

with the scale parameter $\gamma$.

Assuming that secondary effects (attachment, photoproduction and space charge) can be neglected, electron multiplication in gas can be expressed in terms of the first Townsend coefficient $\alpha$. With the additional simplifying assumptions of a uniform field across the THGEM holes and full electron transfer efficiency from one THGEM to the successive one, the gain can be basically written as:

$$G = \exp(\alpha\, d) \sim \exp(V\, \#_{THGEM}) \qquad \text{Eq. 2}$$

In the above equation, d is the total length of the multiplication region, which in turn is proportional to the number of cascade THGEM elements (d ~ #$_{THGEM}$).

The generalized form of α as a function of the reduced field is quite complex. Nevertheless, in He and at relative low reduced electric field (<20 Volt/cm/torr), one may assume that the first Townsend coefficient is linearly proportional to the electric field [25], and thus to the voltage bias (V) applied to each single THGEM multiplier (α ~ V). Combining the two equations, one obtains

$$\frac{\gamma}{p} \sim \text{const. } \#_{THGEM} \quad \text{Eq. 3}$$

The reduced scale parameter (γ/p) depends only on the detector configuration, being solely directly proportional to the number of cascade THGEMs elements (#$_{THGEM}$).

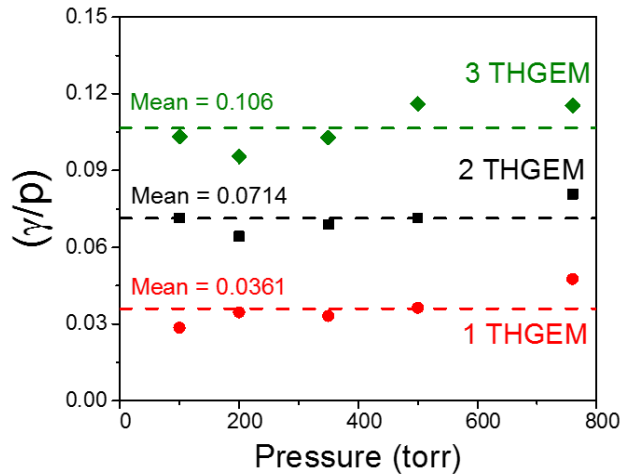

**Figure 6. Reduced scale parameter (γ/p) as a function of the Helium pressure for the three different detector configuration (single-, double- and triple cascade THGEM). The reduced scale parameters were computed by fitting the gain curves of Figure 5 with an exponential function model.**

Figure 6 shows the reduced scale parameter computed from the experimental gain curves of Figure 5 for each detector configuration. The data in Figure 6 has a large uncertainty (of the order of ±20%, not shown in the graphs), mostly attributed to the large error in measuring the gain curve in current mode (particularly in defining the normalization current $I_{IN}$) as well as variations of operational conditions (gas composition, UV-source instabilities, etc.) which may affect the measurements.

In spite of all the simplifying assumptions and a slightly systematic increase of the reduced scale parameter (γ/p) for high pressures, the validation of the above arguments, summarized by Eq. 3, is experimentally significant. The proportional mode of the THGEM detector in helium is preserved for a large pressure range (100-760 torr) and for a wide range of reduced field strength. Nevertheless, a significant deviation from the exponential gain growth has been observed for extremely high reduced bias (>2.5 Volt/torr) and for very low pressure (100 torr) – see for example the gain curve of double-THGEM detector in helium at 100 torr depicted in Figure 7a. This effect can be explained as follows: at low pressure the mean free path of the electrons increases so that a high electric field is needed to reach a certain electron avalanche multiplication. However, an increase of the electric field corresponds to a proportionate expansion of the volume where the avalanche multiplication takes place. Loss of the avalanche confinement is reached at very high field, when the avalanche region extends out of the THGEM holes (Figure 7c), causing the metalized electrode (copper) of the THGEM to be exposed to the radiative emission from the avalanche process. The result is a substantial photo-feedback effect in which the secondary photons impinging on the copper electrode of the THGEM induce secondary photoemission, and consequently generate new avalanches. The secondary photoemissions, triggered by large and extended avalanches, cause the measured single-electron pulse-height spectra (black graph in Figure 7b) to deviate from the typical exponentially decreasing Polya distribution (red graph in Figure 7b).

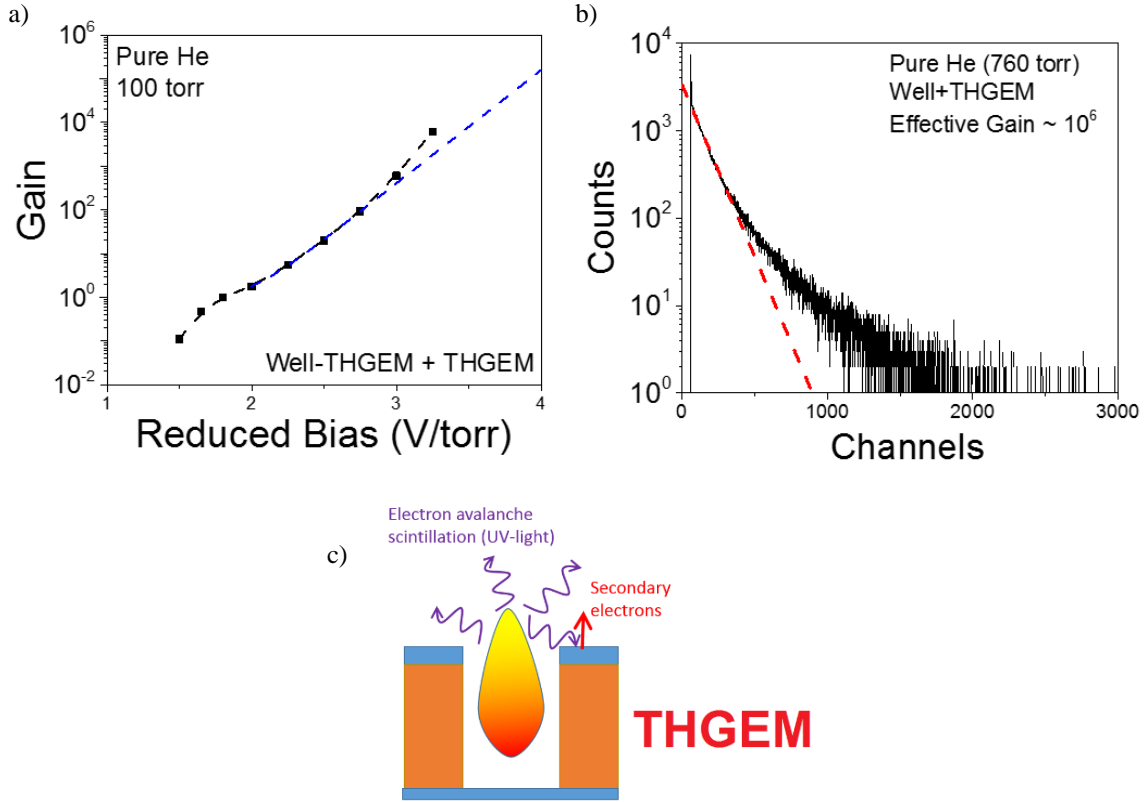

Figure 7. Part a: deviation from the exponential behavior as consequence of avalanche-mediated secondary effects on detector gain curve (part a) and on the single-electron spectrum (part b), for THGEM operates in pure helium at 100 torr. Secondary effects may be explained as loss of electron avalanche confinement that results in photon-feedback (part c).

## 3.2 Long-term gain stability

The gain stability over time of hole-type multipliers is known to be related to several factors [26], [27], including radiation-induced charging up and polarization of the insulator substrate (Kapton, FR4, etc.), and time-dependent variation of the chemical composition of the filling gas (outgassing from the detector components, water-content, residual gases). The detector gain variation and the stabilization time is also expected to be related to the pressure of the gas filling and to the electric dipole field strength across the THGEM hole (detector gain). In this section we present the investigation of long-term gain variations for THGEM detectors, measured for the first time in low-pressure helium.

Figure 8 depicts measurements performed with a triple-cascade THGEM detector, in two different experimental procedures. In the first experiment (Figure 8a), the data were acquired immediately after evacuating and refilling the detector vessel; in the second one (Figure 8b), after several hours of constant gas flow at a rate of 20-30 sccm. In both measurements, a voltage difference of around 290 Volt was symmetrically applied across the THGEM multipliers, while the detector was operating at atmospheric pressure (760 torr). The detector gain was monitored by recording the avalanche current on the anode readout.

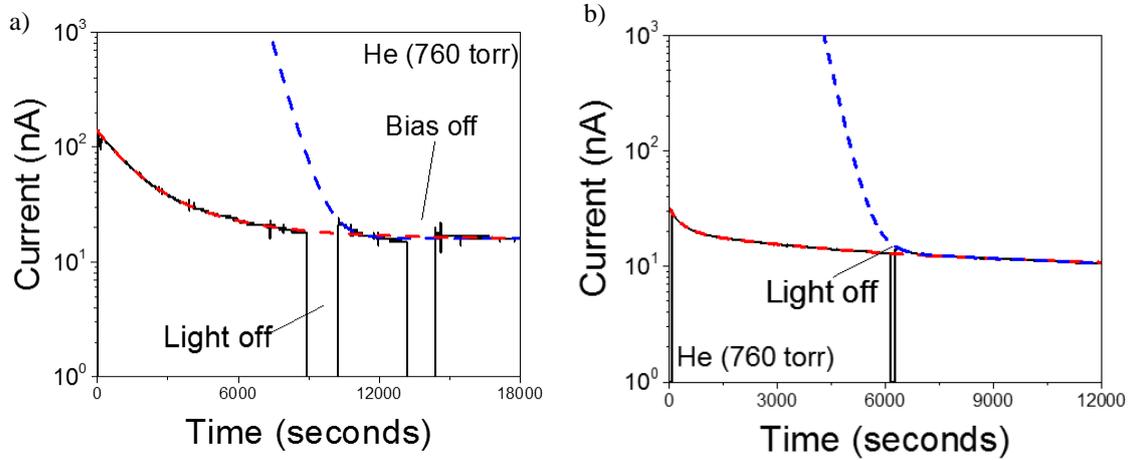

**Figure 8.** Time-variation of the anode current measured from a triple-cascade THGEM detector operating at a gain of around 300 in Helium at a pressure of 760 torr. In part a) the measurement started just after evacuating and then refilling the detector vessel, while in part b) the measurement were taken after several hours of constant gas flow, at a rate of 20-30 sccm.

From the comparison of the graphs in Figure 8, it is evident that a large initial gain variation (factor ~ 10) observed in the first experimental arrangement (Figure 8a) is mainly dominated by variation of the amount of impurities in the filling gas as a result of outgassing from the various detector components. This process (red-graphs) has a relative long time-scale; a constant gain is achieved when the impurities in the filling gas reach a stable concentration level after several hours of constant gas flow. The small initial gain variation (factor ~2) shown in Figure 8b may be attributed to charging up and polarization of the THGEM insulator substrate, which takes place within the first hour from the start of the measurement. To test the impact of the charging up effect on gain stability, the UV-light illuminating the first THGEM cascade was temporarily switched off ("light off" in Figure 8): in this case, the gain variation were found to be proportional to the time during which the detector was not irradiated. The temporal change is characterized by a well-defined time constant (blue graphs). It is interesting to note that, though at a different intensity and with a different time scale, similar gain variations of thick-GEM like structures due to charging up effect and outgassing were also reported by other group in pure noble gas (see for example [28], [29]).

The gain stability was also investigated at a lower pressure (300 torr) using the same methodology as for the data depicted in Figure 8a; i.e. the anode current was recorded just after the vessel was evacuated and refilled at the desired pressure (Figure 9). In this case, the initial variation of the detector gain over time is even larger (factor >10) and the stabilization time is definitely longer, exceeding three hours. This result may be explained assuming that at lower pressure the outgassing components become more relevant and a longer time is needed before a significant amount of the impurities are removed.

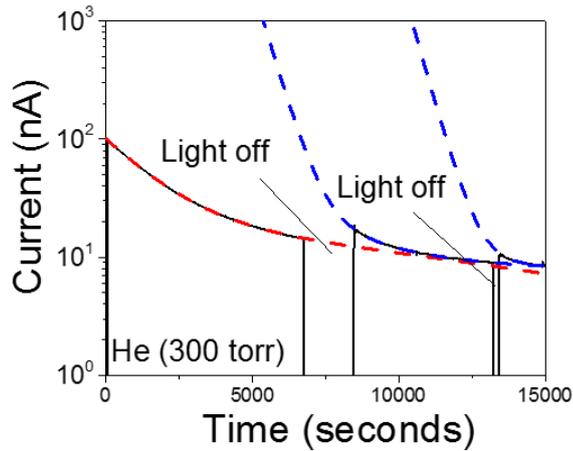

**Figure 9.** Time-variation of the anode current measured from a triple-cascade THGEM detector operating at a gain of around 400, in Helium at a pressure of 300 torr.

### 3.3 Alpha tracks and energy resolution

Figure 10 shows examples of signals from the 25 strips of the segmented anode recorded by the T2K electronics. The full set of data was collected from 10 trigger events, corresponding to 10 alpha particles crossing the drift volume of the detector, operated at a gain of ~300 in 500 torr He. As the area of the THGEM is smaller than the total area covered by the segmented readout, the two edge strips (1 and 25) have a smaller charge collection (around half), and thus a smaller signal pulse-height compared to the inner strips.

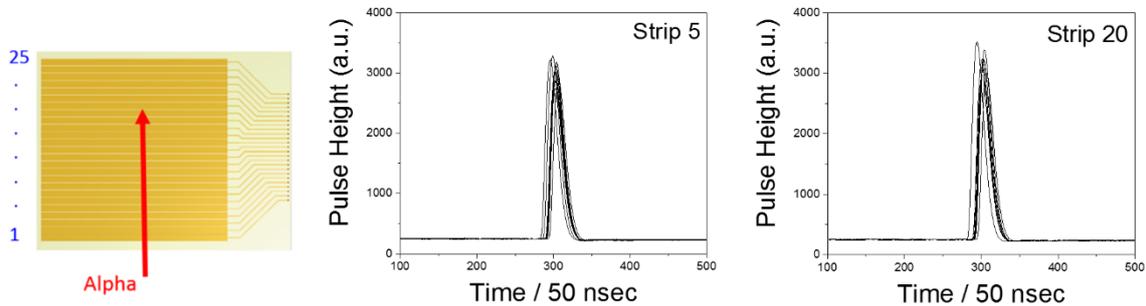

**Figure 10.** Examples of signals from 10 alpha track recorded by the T2K electronics on strip 5 and strip 20. The data were taken in 500 torr helium, at a detector gain of around 300.

In addition to the aforementioned decreased signal pulse-height on the edge strips, we observed a significant effective gain inhomogeneity across the active area of the detector. Figure 11 illustrates a comparison between the experimental pulse-height profiles (part b) and the expected energy loss distributions (part a) for three different gas pressures (760, 500 and 350 torr), computed by Monte Carlo simulations with the SRIM package. The charge amplitudes of each strip, plotted in Figure 11b, are computed by averaging 1000 alpha particle events and scaling them by the given effective gain (shown in the graphs). Beside rather large systematic fluctuations in all of the experimental pulse-height profiles, the deviation from the theoretical energy-loss is quite significant. For instance, at low pressures (500 and 350 torr), where a rather uniform energy loss across the full length of the detector is expected, the spread of the measured pulse-height distributions are of the order of 20% (FWHM). Gain fluctuations may be the result of geometry imperfections of the THGEM multiplier originating from the mechanical drilling process (such as variation of the hole diameter and of the insulator substrate thickness) from the chemical etching (resulting in different widths of the copper rim around the holes), from mechanical defects of the detector assembly (support structures and readout board), or from intrinsic gain variations between channels of the front-end electronics.

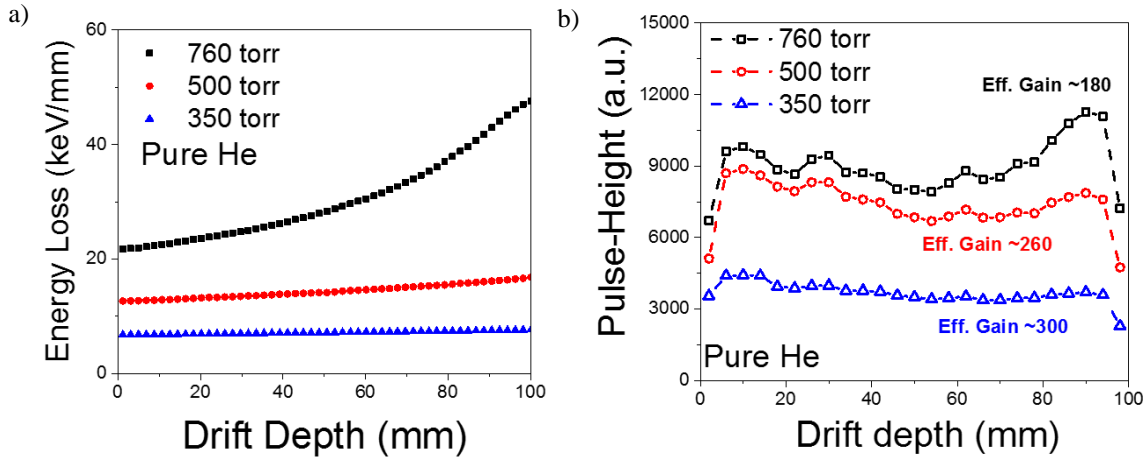

**Figure 11.** Energy loss computed by SRIM simulation (part a) compared to measured pulse-height distributions along the 25 strips (part b) for 5.5 MeV alpha particles crossing the active area of the detector, at various pressure (350-760 torr). The pulse-height distribution in part b) were scaled for the THGEM detector gain.

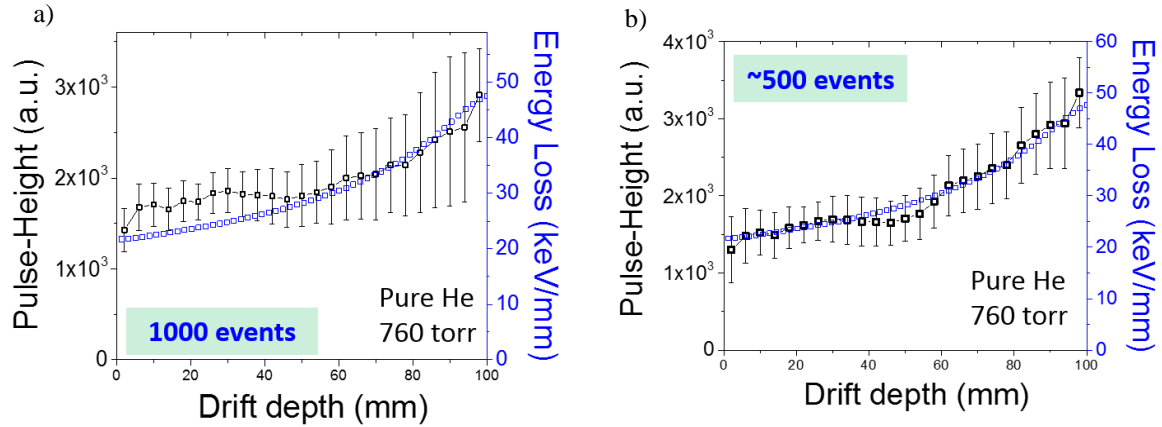

**Figure 12.** Average pulse-height profiles, after strip-to-strip gain correction, of 5.5 MeV alpha tracks in He at 760 torr for the all of the measured trigger events (part a) and for selected zero-angle trigger events (part b).

Correction of the detector gain inhomogeneity is usually performed as a post-processing analysis, for instance using flat-field correction techniques [30]. With this approach, the pulse-height profile of an alpha track measured in He at 350 torr, resulting from a uniform energy loss, was taken as a calibration for correcting the profile measured at 760 torr. Figure 12a depicts the 760 torr pulse-height profile after the strip-to-strip gain variation was corrected, while Figure 12b shows the pulse-height distribution after discarding all tracks of particles emitted from the source at non-zero angles; these alpha particles are responsible for the exceeding signal pulse-heights at the entrance of the detector active area, clearly noticeable in Figure 12a. The non-zero-angle discrimination of the alpha tracks were performed by selecting all events in which the signals recorded on strip 24 and strip 25 were above a well-defined threshold; this ensures that alpha particles were indeed crossing the full drift gap of the detector. The experimental data of Figure 12 is compared to the expected energy loss distribution (blue graphs), and are well reproduced by the corrected pulse-height profile of Figure 12b.

As the range of 5.5 MeV alpha particles in pure He is much longer than the length of the detector active area for any of the pressure values consider in this study, it was not possible to derive an absolute value of the energy resolution of the full track. Instead, we computed the relative energy resolution based on the energy deposited on each strip and on the full detector active area from the uniform energy loss when the THGEM detector was operated at 350 torr. At this pressure the alpha particles deposit an energy of 29 keV on each strip and a total energy of 690 keV in the full detector. As shown in Figure 13a, an average energy resolution of 18.5% was attained on various readout strips. Examples of spectra recorded on three different strips are shown in Figure 13b.

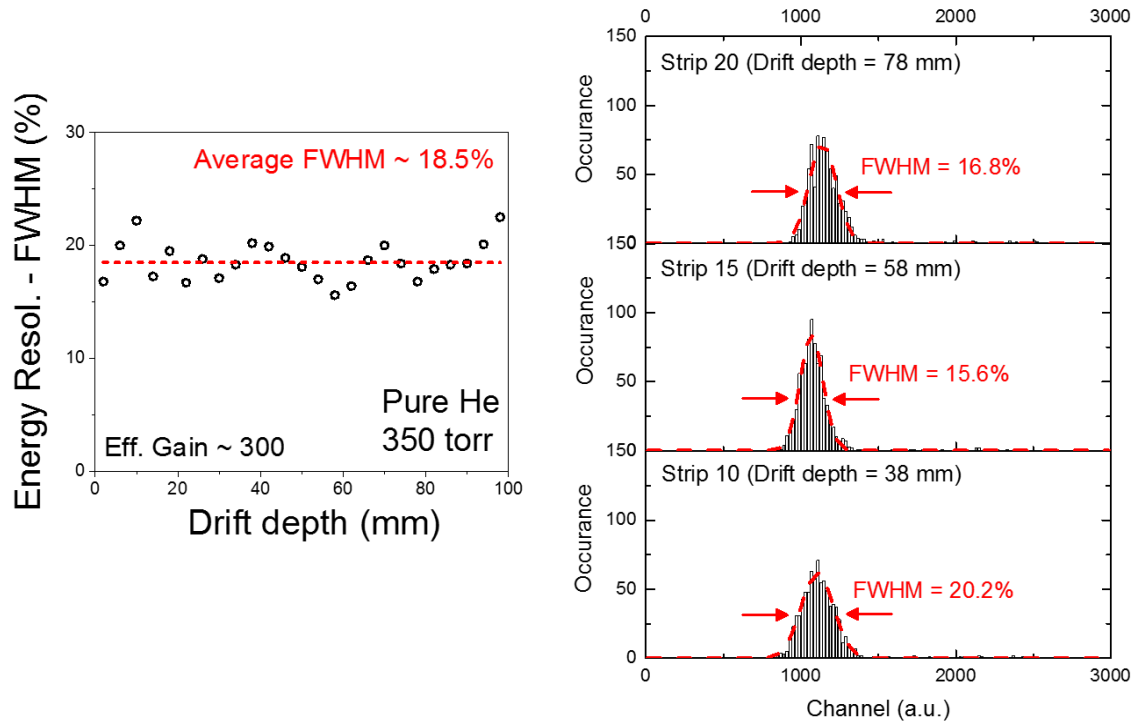

**Figure 13. Energy resolution (FWHM) from tracks of 5.5 MeV alpha particles measured on the 25 strips readout at 350 torr (part a). Examples of spectra measured with the T2K electronics in three different strips (10, 15 and 20).**

The spectrum of the energy deposited by the alpha particle on the entire active area of the detector is depicted in Figure 14; it was computed by summing up the charge collected by the anode strips in each events. An energy resolution of 6.65% (FWHM) was obtained for 690 keV, while the expected value for 5.5 MeV can be estimated as $6.65\% \times \sqrt{0.69/5.5} \approx 2.4\%$ (FWHM). Better energy resolution may be achieved by optimizing the field in the drift gap as well as between the THGEM elements.

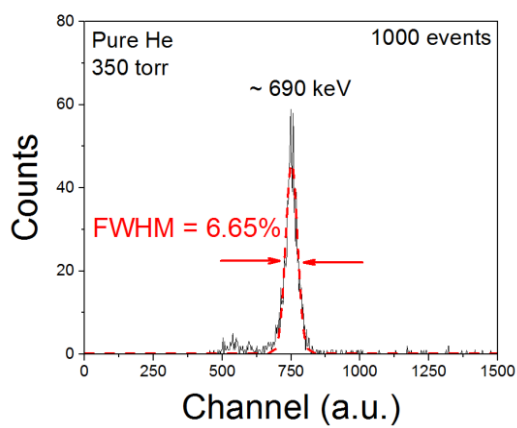

Figure 14. Measured energy spectrum deposited by 5.5 MeV alpha in the detector active area (~690 keV) in He at 350 torr.

## 4. Summary and conclusion

Within this work the operational characteristics and properties of the THGEM-based detectors have been investigated in pure helium at low pressure, from 760 torr down to 100 torr. Due to the extended thickness of the multiplication region within the THGEM holes, several times larger than the mean free path of the avalanche electrons even at low pressure, stable operation conditions at high effective gain could be attained in pure He. For instance, with three THGEM elements in cascade spark-free effective gains of $10^4$ and over $10^7$ (with single photoelectrons) were obtained at 100 and 760 torr, respectively. The high maximum achievable gain in He, reached at low operational voltage, may be explained as the result of a strong avalanche confinement within the THGEM holes, which limits photo-mediated secondary effects, as well as a possible Penning effect due to an unavoidable small amount of residual impurities (mainly $N_2$). Obvious secondary effects (in the form of photon-feedback), as a consequence of loss of avalanche confinement, were observed at very high reduced voltage bias and only for very low pressure.

A large variation of the effective gain has been observed as a function of the chemical composition of the filling gas, so that several hours are needed for the stabilization of the gain value due to initial outgassing from detector components. Minor effects as a consequence of radiation-induced charging up were observed.

Despite a large gain inhomogeneity across the detector area, an energy resolution of 18.5% (FWHM) in single strip (4 mm wide) were attained from a uniform energy loss (7.1 keV/mm) of 5.5 MeV alpha particle at 350 torr. From the analysis of the spectrum of energy deposited on the full active area (690 keV) and summing up the charge in each strip, we computed an energy resolution 6.65%; this allow us to estimate an energy resolution of 2.4% for the full 5.5 MeV alpha particle track.

The results obtained in the present study demonstrate the excellent performance of THGEM multiplier operating in low-pressure helium, which are of relevance for future AT-TPC applications. Further investigations of the properties of THGEM-based TPCs with active target, in respect to spatial and angular resolution, ion-feedback flow reduction and operation in pure Hydrogen and pure Deuterium are in progress.

## Acknowledgments

The authors would like to thank the AT-TPC group of NSCL (MSU), in particular D. Bazin, W. Mittig and S. Beceiro Novo, for the support and fruitful discussions. In addition, the authors would also like to thank T. Ahn (University of Notre Dame) for his assistance in operating the T2K electronics.

## References

[1] T. Francke and V. Peskov, *Innovative applications and developments of micro-pattern gaseous detectors*. Hershey, PA: Information Science Reference, 2014.

[2] F. Sauli, "GEM: A new concept for electron amplification in gas detectors," *Nucl. Instrum. Methods Phys. Res. Sect. Accel. Spectrometers Detect. Assoc. Equip.*, vol. 386, no. 2–3, pp. 531–534, Feb. 1997.

[3] Y. Giomataris, P. Rebourgeard, J. P. Robert, and G. Charpak, "MICROMEGAS: a high-granularity position-sensitive gaseous detector for high particle-flux environments," *Nucl. Instrum. Methods Phys. Res. Sect. Accel. Spectrometers Detect. Assoc. Equip.*, vol. 376, no. 1, pp. 29–35, Jun. 1996.

[4] D. Suzuki, A. Shore, W. Mittig, J. J. Kolata, D. Bazin, M. Ford, T. Ahn, F. D. Becchetti, S. Beceiro Novo, D. Ben Ali, B. Bucher, J. Browne, X. Fang, M. Febbraro, A. Fritsch, E. Galyaev, A. M. Howard, N. Keeley, W. G. Lynch, M. Ojaruega, A. L. Roberts, and X. D. Tang, "Resonant α scattering of 6He: Limits of clustering in 10Be," *Phys. Rev. C*, vol. 87, no. 5, p. 054301, May 2013.

[5] D. Suzuki, D. Bazin, W. Mittig, W. G. Lynch, C. Hewko, A. Roux, D. Ben Ali, J. Browne, E. Galyaev, M. Ford, A. Fritsch, J. Gilbert, F. Montes, A. Shore, G. Westfall, and J. Yurkon, "Test of a micromegas detector with helium-based gas mixtures for active target time projection chambers utilizing radioactive isotope beams," *Nucl. Instrum. Methods Phys. Res. Sect. Accel. Spectrometers Detect. Assoc. Equip.*, vol. 660, no. 1, pp. 64–68, Dec. 2011.

[6] J. Pancin, S. Damoy, D. Perez Loureiro, V. Chambert, F. Dorangeville, F. Druillole, G. F. Grinyer, A. Lermitage, A. Maroni, G. Noël, C. Porte, T. Roger, P. Rosier, and L. Suen, "Tests of Micro-Pattern Gaseous Detectors for active target time projection chambers in nuclear physics," *Nucl. Instrum. Methods Phys. Res. Sect. Accel. Spectrometers Detect. Assoc. Equip.*, vol. 735, pp. 532–540, Jan. 2014.

[7] H. Ishiyama, K. Yamaguchi, Y. Mizoi, Y. X. Watanabe, S. K. Das, T. Hashimoto, H. Miyatake, Y. Hirayama, N. Imai, M. Oyaizu, S. C. Jeong, T. Fukuda, S. Mitsuoka, H. Makii, and T. K. Sato, "GEM-MSTPC: An active-target type detector in low-pressure He/CO2 mixed gas," *J. Instrum.*, vol. 7, no. 03, p. C03036, Mar. 2012.

[8] K. Yamaguchi, H. Ishiyama, M. H. Tanaka, Y. X. Watanabe, H. Miyatake, Y. Hirayama, N. Imai, H. Makii, Y. Fuchi, S. C. Jeong, T. Nomura, Y. Mizoi, S. K. Das, T. Fukuda, T. Hashimoto, and I. Arai, "Development of the GEM-MSTPC for measurements of low-energy nuclear reactions," *Nucl. Instrum. Methods Phys. Res. Sect. Accel. Spectrometers Detect. Assoc. Equip.*, vol. 623, no. 1, pp. 135–137, Nov. 2010.

[9] A. Bressan, A. Buzulutskov, L. Ropelewski, F. Sauli, and L. Shekhtman, "High gain operation of GEM in pure argon," *Nucl. Instrum. Methods Phys. Res. Sect. Accel. Spectrometers Detect. Assoc. Equip.*, vol. 423, no. 1, pp. 119–124, Feb. 1999.

[10] A. Bondar, A. Buzulutskov, and L. Shekhtman, "High pressure operation of the triple-GEM detector in pure Ne, Ar and Xe," *Nucl. Instrum. Methods Phys. Res. Sect. Accel. Spectrometers Detect. Assoc. Equip.*, vol. 481, no. 1–3, pp. 200–203, Apr. 2002.

[11] J. Miyamoto, A. Breskin, and V. Peskov, "Gain limits of a Thick GEM in high-purity Ne, Ar and Xe," *J. Instrum.*, vol. 5, no. 05, p. P05008, May 2010.

[12] M. Cortesi, V. Peskov, G. Bartesaghi, J. Miyamoto, S. Cohen, R. Chechik, J. M. Maia, J. M. F. dos Santos, G. Gambarini, V. Dangendorf, and A. Breskin, "THGEM operation in Ne and Ne/CH4," *J. Instrum.*, vol. 4, no. 08, p. P08001, Aug. 2009.

[13] R. Alon, J. Miyamoto, M. Cortesi, A. Breskin, R. Chechik, I. Carne, J. M. Maia, J. M. F. dos Santos, M. Gai, D. McKinsey, and V. Dangendorf, "Operation of a Thick Gas Electron Multiplier (THGEM) in Ar, Xe and Ar-Xe," *J. Instrum.*, vol. 3, no. 01, p. P01005, Jan. 2008.

[14] A. Buzulutskov, J. Dodd, R. Galea, Y. Ju, M. Leltchouk, P. Rehak, V. Tcherniatine, W. J. Willis, A. Bondar, D. Pavlyuchenko, R. Snopkov, and Y. Tikhonov, "GEM operation in helium and neon at low temperatures," *Nucl. Instrum. Methods Phys. Res. Sect. Accel. Spectrometers Detect. Assoc. Equip.*, vol. 548, no. 3, pp. 487–498, Aug. 2005.

[15] C. S. Lee, S. Ota, H. Tokieda, R. Kojima, Y. N. Watanabe, and T. Uesaka, "Properties of thick GEM in low-pressure deuterium," *J. Instrum.*, vol. 9, no. 05, p. C05014, May 2014.

[16] A. Breskin, R. Alon, M. Cortesi, R. Chechik, J. Miyamoto, V. Dangendorf, J. M. Maia, and J. M. F. Dos Santos, "A concise review on THGEM detectors," *Nucl. Instrum. Methods Phys. Res. Sect. Accel. Spectrometers Detect. Assoc. Equip.*, vol. 598, no. 1, pp. 107–111, Jan. 2009.

[17] C. Shalem, R. Chechik, A. Breskin, and K. Michaeli, "Advances in Thick GEM-like gaseous electron multipliers—Part I: atmospheric pressure operation," *Nucl. Instrum. Methods Phys. Res. Sect. Accel. Spectrometers Detect. Assoc. Equip.*, vol. 558, no. 2, pp. 475–489, Mar. 2006.

[18] R. Chechik, A. Breskin, C. Shalem, and D. Mörmann, "Thick GEM-like hole multipliers: properties and possible applications," *Nucl. Instrum. Methods Phys. Res. Sect. Accel. Spectrometers Detect. Assoc. Equip.*, vol. 535, no. 1–2, pp. 303–308, Dec. 2004.

[19] C. K. Shalem, R. Chechik, A. Breskin, K. Michaeli, and N. Ben-Haim, "Advances in thick GEM-like gaseous electron multipliers Part II: Low-pressure operation," *Nucl. Instrum. Methods Phys. Res. Sect. Accel. Spectrometers Detect. Assoc. Equip.*, vol. 558, no. 2, pp. 468–474, Mar. 2006.

[20] C. Angulo, "Experimental Tools for Nuclear Astrophysics," in *The Euroschool Lectures on Physics with Exotic Beams, Vol. III*, J. S. Al-Khalili and E. Roeckl, Eds. Springer Berlin Heidelberg, 2009, pp. 253–282.

[21] Yamaguchi, D. Kahl, T. Nakao, Y. Wakabayashi, S. Kubono, T. Hashimoto, S. Hayakawa, T. Kawabata, N. Iwasa, T. Teranishi, Y. K. Kwon, P. S. Lee, D. N. Binh, L. H. Khiem, and N. G. Duy, "Studies on alpha-induced astrophysical reactions using the low-energy RI beam separator CRIB," *EPJ Web Conf.*, vol. 66, p. 4, 2014.

[22] L. Arazi, H. N. da Luz, D. Freytag, M. Pitt, C. D. R. Azevedo, A. Rubin, M. Cortesi, D. S. Covita, C. a. B. Oliveira, E. Oliveri, R. Herbst, S. Park, J. Yu, R. Chechik, J. M. F. dos Santos, M. Breidenbach, G. Haller, A. White, J. F. C. A. Veloso, and A. Breskin, "THGEM-based detectors for sampling elements in DHCAL: laboratory and beam evaluation," *J. Instrum.*, vol. 7, no. 05, p. C05011, May 2012.

[23] S. Bressler, L. Arazi, L. Moleri, M. Pitt, A. Rubin, and A. Breskin, "Recent advances with THGEM detectors," *J. Instrum.*, vol. 8, no. 12, p. C12012, Dec. 2013.

[24] P. Baron, D. Calvet, E. Delagnes, X. De La Broise, A. Delbart, F. Druillole, E. Mazzucato, E. Monmarthe, F. Pierre, and M. Zito, "AFTER, an ASIC for the Readout of the Large T2K Time Projection Chambers," *IEEE Trans. Nucl. Sci.*, vol. 55, no. 3, pp. 1744–1752, Jun. 2008.

[25] L. M. Chanin and G. D. Rork, "Experimental Determinations of the First Townsend Ionization Coefficient in Helium," *Phys. Rev.*, vol. 133, no. 4A, pp. 1005–A1009, Feb. 1964.

[26] R. Chechik, M. Cortesi, A. Breskin, D. Vartsky, D. Bar, and V. Dangendorf, "Thick GEM-like (THGEM) Detectors and Their Possible Applications," *SNIC Symp. Stanf. Calif. USA*, pp. 25–28, Apr. 2006.

[27] B. Azmoun, W. Anderson, D. Crary, J. Durham, T. Hemmick, J. Kamin, G. Karagiorgi, K. Kearney, G. Keeler, E. Kornacki, P. Lynch, R. Majka, M. Rumore, F. Simon, J. Sinsheimer, N. Smirnov, B. Surrow, and C. Woody, "A Study of Gain Stability and Charging Effects in GEM Foils," in *IEEE Nuclear Science Symposium Conference Record, 2006*, 2006, vol. 6, pp. 3847–3851.

[28] C. Cantini, L. Epprecht, A. Gendotti, S. Horikawa, L. Periale, S. Murphy, G. Natterer, C. Regenfus, F. Resnati, F. Sergiampietri, A. Rubbia, T. Viant, and S. Wu, "Performance study of the effective gain of the double phase liquid Argon LEM Time Projection Chamber," *ArXiv14124402 Phys.*, Dec. 2014.

[29] C. Cantini, L. Epprecht, A. Gendotti, S. Horikawa, S. Murphy, G. Natterer, L. Periale, F. Resnati, A. Rubbia, F. Sergiampietri, T. Viant, and S. Wu, "Long-term operation of a double phase LAr LEM Time Projection Chamber with a simplified anode and extraction-grid design," *J. Instrum.*, vol. 9, no. 03, p. P03017, Mar. 2014.

[30] M. Cortesi, R. Zboray, A. Kaestner, and H.-M. Prasser, "Development of a cold-neutron imaging detector based on thick gaseous electron multiplier," *Rev. Sci. Instrum.*, vol. 84, no. 2, p. 023305, Feb. 2013.